\documentstyle[referee]{mn}
\input psfig.tex
\def\lsim{\lower.5ex\hbox{$\; \buildrel < \over \sim \;$}}
\def\gsim{\lower.5ex\hbox{$\; \buildrel > \over \sim \;$}}
\def \simeq{\lower.3ex\hbox{$\; \buildrel \sim \over - \;$}}
\def\ch{\lower-0.55ex\hbox{--}\kern-0.55em{\lower0.15ex\hbox{$h$}}}
\def\lh{\lower-0.55ex\hbox{--}\kern-0.55em{\lower0.15ex\hbox{$\lambda$}}}
\def\be{\begin{equation}}
\def\ee{\end{equation}}
\def\eng{{\cal E}}
\def\vel{\vartheta}
\def\ga{\gamma}
\def\la{\lambda}
\def\md{\dot{\cal M}}
\def\sig{\Sigma}
\def\al{\alpha}

\def\bt{\beta}
\def\btp{\beta^{'}}
\def\dl{\delta}

\newif\ifAMStwofonts
\ifoldfss
  \ifCUPmtlplainloaded \else
    \NewTextAlphabet{textbfit} {cmbxti10} {}
    \NewTextAlphabet{textbfss} {cmssbx10} {}
    \NewMathAlphabet{mathbfit} {cmbxti10} {} 
    \NewMathAlphabet{mathbfss} {cmssbx10} {} 
  \fi
  \ifAMStwofonts
    \ifCUPmtlplainloaded \else
      \NewSymbolFont{upmath} {eurm10}
      \NewSymbolFont{AMSa} {msam10}
      \NewMathSymbol{\upi}     {0}{upmath}{19}
      \NewMathSymbol{\umu}     {0}{upmath}{16}
      \NewMathSymbol{\upartial}{0}{upmath}{40}
      \NewMathSymbol{\leqslant}{3}{AMSa}{36}
      \NewMathSymbol{\geqslant}{3}{AMSa}{3E}

    \fi
  \fi
\fi 
\ifnfssone
  \newmathalphabet{\mathit}
  \addtoversion{normal}{\mathit}{cmr}{m}{it}
  \addtoversion{bold}{\mathit}{cmr}{bx}{it}
  \newmathalphabet{\mathbfit} 
  \addtoversion{normal}{\mathbfit}{cmr}{bx}{it}
  \addtoversion{bold}{\mathbfit}{cmr}{bx}{it}
  \newmathalphabet{\mathbfss} 
  \addtoversion{normal}{\mathbfss}{cmss}{bx}{n}
  \addtoversion{bold}{\mathbfss}{cmss}{bx}{n}
  \ifAMStwofonts
    \ifCUPmtlplainloaded \else
      \UseAMStwoboldmath
      \makeatletter
      \new@mathgroup\upmath@group
      \define@mathgroup\mv@normal\upmath@group{eur}{m}{n}
      \define@mathgroup\mv@bold\upmath@group{eur}{b}{n}
      \edef\UPM{\hexnumber\upmath@group}
      \new@mathgroup\amsa@group
      \define@mathgroup\mv@normal\amsa@group{msa}{m}{n}
      \define@mathgroup\mv@bold\amsa@group{msa}{m}{n}
      \edef\AMSa{\hexnumber\amsa@group}
      \makeatother
      \mathchardef\upi="0\UPM19
      \mathchardef\umu="0\UPM16
      \mathchardef\upartial="0\UPM40
      \mathchardef\leqslant="3\AMSa36
      \mathchardef\geqslant="3\AMSa3E
    \fi
  \fi
\fi 

\ifnfsstwo
  \DeclareMathAlphabet{\mathbfit}{OT1}{cmr}{bx}{it}
  \SetMathAlphabet\mathbfit{bold}{OT1}{cmr}{bx}{it}
  \DeclareMathAlphabet{\mathbfss}{OT1}{cmss}{bx}{n}
  \SetMathAlphabet\mathbfss{bold}{OT1}{cmss}{bx}{n}
  \ifAMStwofonts
    \ifCUPmtlplainloaded \else
      \DeclareSymbolFont{UPM}{U}{eur}{m}{n}
      \SetSymbolFont{UPM}{bold}{U}{eur}{b}{n}
      \DeclareSymbolFont{AMSa}{U}{msa}{m}{n}
      \DeclareMathSymbol{\upi}{0}{UPM}{"19}
      \DeclareMathSymbol{\umu}{0}{UPM}{"16}
      \DeclareMathSymbol{\upartial}{0}{UPM}{"40}
      \DeclareMathSymbol{\leqslant}{3}{AMSa}{"36}
      \DeclareMathSymbol{\geqslant}{3}{AMSa}{"3E}
    \fi
  \fi
\fi 

\ifCUPmtlplainloaded \else
  \ifAMStwofonts \else 
    \def\upi{\pi}
    \def\umu{\mu}
    \def\upartial{\partial}
  \fi
\fi
\title{Model Dependence of Transonic Properties of Accretion Flows Around Black Holes} 
\author[Sandip K. Chakrabarti, Santabrata Das]
       {Sandip K. Chakrabarti$^{1,2}$, Santabrata Das$^1$\\
$^1$ S.N. Bose National Centre for Basic Sciences,\\
JD-Block, Sector III, Salt Lake, Kolkata 700098, India;\\
$^2$ Centre for Space Physics, 114/v/1A Raja S.C. Mullick Rd., Kolkata 700047, India\\
chakraba@boson.bose.res.in,sbdas@boson.bose.res.in}

\date{Accepted .
      Received ;
      in original form }
\pubyear{2000}
\begin{document}

\maketitle

\begin{abstract}
We analytically study how the behaviour of accretion flows changes
when the flow model is varied. We study the transonic properties of the conical
flow, a flow of constant height and a flow in vertical equilibrium
and show that all these models are basically identical provided the
polytropic constant is suitably changed from one model to another.
We show that this behaviour is extendible even when standing shocks
are produced in the flow. The parameter space where shocks are produced remain 
roughly identical in all these models when the same transformation among 
the polytropic indices is used. We present applications of these findings.
\end{abstract}


\section{Introduction}

Fully self-consistent study of any astrophysical system is generally
prohibitive. Very often, for simplicity, it is necessary to construct models which 
have all the salient features of the original problem. However these models 
need not be unique. In the present paper we make a pedagogical review of three different 
models of rotating accretion flows and show that even though they are based on 
fundamentally different assumptions, they have identical physical properties.
What is more, results of one model could be obtained from the other by changing
a {\it physical} parameter, namely, the polytropic constant. In other words, all these
models are {\it identical}.

Accretion disk physics has undergone major changes in the last fifty years. 
Bondi (1952) studied spherical accretion and found the existence of only
one saddle type sonic point in an adiabatic flow. Later, the Keplerian 
disk model of Shakura and Sunyaev (1973) and thick disk model of 
Paczy\'nski and Wiita (1980) the disk solutions became more realistic,
though none of them was transonic, i.e., none was passing through any sonic point.
Meanwhile, Liang and Thompson (1980) generalized this work for a flow which included
angular momentum and discovered that there could be three sonic points. 
Matsumoto et al. (1984) tried to let the flow pass through the inner sonic point only
and found that flow could pass through nodal type sonic points.

Chakrabarti (1989, 1990; hereafter referred to as C89 and C90 respectively) 
studied transonic properties  of accretion flows which are conical 
in shape in the meridional plane (`Wedge-shaped Flow') 
and also flows which are in vertical equilibrium. Subsequently, Chakrabarti
\& Molteni (1993) studied flows of constant height and also verified 
by time dependent numerical simulations that the flow indeed allows 
standing shocks in it. In a Bondi (1952) flow, to specify a 
solution one requires exactly one parameter, namely the specific 
energy ${\cal E}$ of the flow. This is in turn determined by the 
temperature of the flow at a large distance. In an inviscid, rotating 
axisymmetric accretion flow, one requires two parameters, namely, 
specific energy ${\cal E}$ and specific angular momentum $\lambda$. 
Once they are specified, all the crucial properties of the flow, 
namely the locations of the sonic points, the shocks, as well as the 
complete global solution are determined. C89 numerically studied 
the properties of the parameter space rather extensively, and divided 
the parameter space in terms of whether standing shocks can form or not. In 
the present paper, we compare these models completely analytically 
and show, very interestingly, that one could easily `map' one 
model onto another by suitably changing the polytropic index
of the flow. In other words, we show that these models are roughly 
identical to one another as far as the transonic properties go.

In the next Section, we present a set of equations which govern
the steady state flow in all the three models. In \S 3, we
present the sonic point analysis and provide 
the expressions for the energy of the flow in terms of the
sonic points. We observe that these expressions are identical 
provided there is a unique relation among the polytropic indices
of these model flows. In \S 4, we compare shock locations 
in all the three models. We also compare the parameter 
space which allows shock formation in these models 
with the regions obtained using purely numerical methods. 
In \S 5, we show that in fact if the relations between 
the polytropic indices are used, the shock locations in all 
these models are also roughly identical. Consequently, 
the apparently disjoint parameter spaces drawn with the same polytropic 
index overlap almost completely when the aforementioned 
relations among polytropic indices is used. This remarkable
behaviour shows underlying unity in these apparently diverse
models. Finally, in \S 6, we draw our conclusions.

\section{Model Equations}

As discussed in the Introduction, we shall be concerned about three 
axisymmetric and inviscid models: (a) {\it Model H}: the flow has a constant height 
everywhere; (b) {\it Model C}: the flow cross-section in meridional plane is conical in shape and
(c) {\it  Model V}: the flow is in equilibrium in the transverse direction.
Figure 1 shows a cartoon diagram of these three models. Filled circle
at the centre correspond to the black hole. Region of the disk shaded in light corresponds
to the pre-shock flow while the region shaded in dark corresponds to  the post-shock
flow. We also assume that the distances are measured in units of $r_S=2GM_{BH}/c^2$, 
where $G$ is the gravitational constant, $c$ is the velocity of light, 
$M_{BH}$ is the mass of the black hole. Velocities and 
angular momenta are measured in units of $c$ and $c r_S = 2GM_{BH}/c$
respectively. In all the three models, the dimensionless 
energy conservation law can be written as,
\be
{\cal E} = {\frac{\vel_{e}^{2}}{2}}+{\frac {a_{e}^{2}}{\ga - 1}}
+{\frac {\la^{2}}{2x^{2}}}+g(x)
\ee
where, $g(x)$ is the pseudo-Newtonian potential introduced by
Paczy\'nski \& Wiita (1980) and is given by, $g(x) = -{\frac {1}{2(x-1)}}$. 
Here, $\vel_{e}$ and $a_{e}=\sqrt{\gamma p/\rho}$ are the non-dimensional 
radial and the sound velocities respectively, $x$ is the non-dimensional 
radial distance, the subscript $e$ refers to the quantities measured on 
the equatorial plane. The flow has been chosen to be adiabatic with
equation of state, $P=K\rho^\gamma$, where $K$ is a constant
which measures the entropy of the flow, and $\gamma$ is the
polytropic exponent. The energy equation is the integral from of the radial 
momentum balance equation.

\begin {figure}
\vbox{
\vskip 0.0cm
\hskip 0.0cm
\centerline{
\psfig{figure=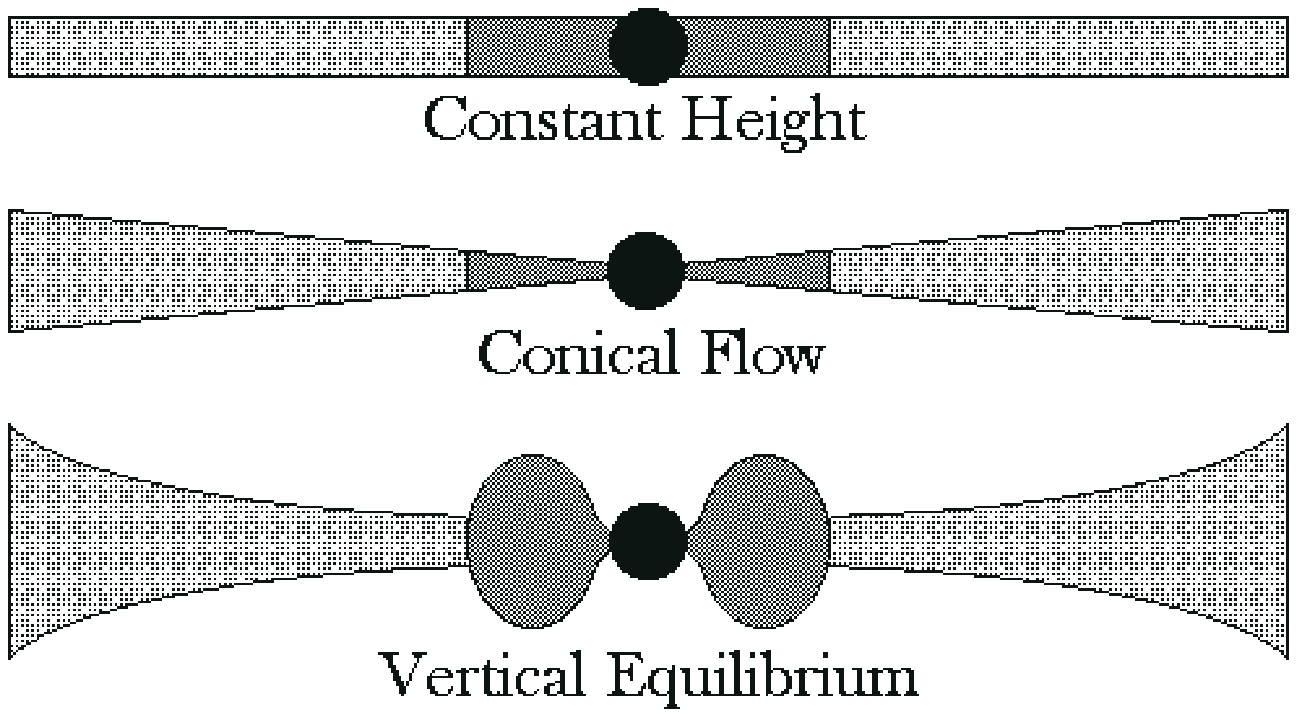,height=10truecm,width=12truecm}}}
\end{figure}
\begin{figure}
\vspace{0.0cm}
\caption[] {Cartoon diagram of three different models discussed in the
text. In constant height flow (H) disk thickness is constant (upper). In a conical flow (C),
the cross-section in the meridional plane is conical (middle). In a vertical equilibrium flow
(V), matter is in locally in vertical equilibrium at every point of the disk (lower).}
\end{figure}

The mass flux conservation equation, which comes directly from the continuity 
equation depends on specific geometry of the models.
Apart from a geometric constant, the conservation equation is given by,
\be
\dot{M} = \vel_{e} \rho a^\zeta  x^{\bt} (x-1)^{\dl} .
\ee
where $\bt$, $\zeta$ and $\dl$ are constants.
For Model V (Chakrabarti, 1989), $\bt = 3/2,~ \zeta=1,~ \dl = 1$.
For Model C (Chakrabarti, 1990), $\bt = 2, \zeta=0, \dl = 0$.
For Model H (Chakrabarti, 1992; Chakrabarti \& Molteni, 1993), $\bt = 1, \zeta=0, \dl = 0$.
Note that since the local disc height $h(x)$
depends on sound speed $h(x)\sim a_e x^{1/2}(x-1)$, so a factor of $a_e^\zeta$ is 
applicable for this Model.

Though it is customary to deal with the conserved mass accretion rate 
of the flow, since we incorporate shock formation where entropy is increased, 
it is more convenient to re-write the mass flux conservation equation in 
terms of $\vel_{e}$ and $a_{e}$ in the following way, 
\be
\md = \vel_{e} a_{e}^{\al}x^{\bt}(x-1)^{\dl} = \vel_e a_e^\al f(x), 
\ee
where, $\al = 2n+\zeta$, $\al=2n$ and $\al=2n$ for Models V, C and H 
respectively and $f(x)=x^\bt (x-1)^\dl$. We shall use the phrase 
``entropy-accretion rate'' for the quantity, ${\dot {\cal M}} = 
\dot {M}K^{n}\ga^{n}$. In a flow without a shock, this quantity 
remains constant, but in presence of a shock it changes because 
of the generation of entropy.

\section{Sonic Point Analysis and Relation Between Models}

Since the flow is expected to be sub-sonic at a long distance and supersonic 
on the horizon, the flow must pass through sonic points. At the sonic point a 
few conditions are to be satisfied. They can be derived in the following way:

First, we differentiate the energy equation and the mass conservation 
equation and eliminate ${da/dx}$ from them to obtain,
\be
\frac {d\vel}{dx}=
\frac{\frac {2na^{2}}{\al}\left[\frac {\btp x-\bt}{x(x-1)}\right]
-\frac {dG}{dx}}{{\left[ \vel -\frac {2na^{2}}{\al \vel}\right]}}
\ee
Here, $G(x)= \frac {\la ^{2}}{2x^{2}}-\frac {1}{2(x-1)}$ 
is the effective potential and $\btp=\beta+\delta$. Since the flow is assumed 
to be smooth everywhere, if at any point of the flow 
denominator vanishes, the numerator must also vanish there. 
The vanishing of the denominator gives,
\be 
\vel^{2}_{c}(x_{c})=\frac {2n}{\al }a^{2}_{c}(x_{c}) .
\ee
The vanishing of the numerator gives,
\be
a^{2}_{c}(x_{c})=\frac {\al(x_{c}-1)}{2nx^{2}_{c}}
\frac {[\la^{2}_{K}(x_{c})-\la^{2}]}{[\btp x_{c}-\bt]}.
\ee
The subscript $c$ denotes quantities at the critical points.
Here, $\la_{K}$ is the Keplerian angular momentum defined as
$\la^{2}_{K}=x^{3}_{c}/{[2(x_{c}-1)^{2}]}$.
It is to be noted that since square of the sound speed (eq. 6) is always 
positive, angular momentum at the sonic point must be 
sub-Keplerian, i.e., $\lambda <\lambda_{\it {K}}$.
When the above expression for the velocity of sound is
inserted in the expression for the specific energy, we get, for the 
{\it {Vertical Equilibrium (V) Model}},
$$
\eng_{\rm V}=\frac {n_{\rm V}+1}{5}\frac {x_c}{(x_c-3/5)(x_c-1)}
-\left[\frac {4(n_{\rm V}+1)}{5}\frac{(x_c-1)}{x_c-3/5}-1\right]
\frac{\la^2}{2x^2_c}-\frac{1}{2(x_c-1)} ,
\eqno{(7a)}
$$
for {\it {the Conical Flow (C) Model}},
$$
\eng_{\rm C}=\frac {2n_{\rm C}+1}{8}\frac{x_c}{(x_c-1)^2}
-\frac{2n_{\rm C}-1}{2}\frac{\la^2}{2x^2_c}-\frac{1}{2(x_c-1)},
\eqno{(7b)}
$$
and {\it {for Constant Height Flow (H) Model}},
$$
\eng_{\rm H}=\frac{2n_{\rm H}+1}{4}\frac{x_c}{(x_c-1)^2}
-2n_{\rm H}\frac{\la^2}{2x^2_c}-\frac{1}{2(x_c-1)} .
\eqno{(7c)}
$$
Here, we have used the subscripts V, C and H under specific 
energy and polytropic index $n$ to denote specific models. 
For a given angular momentum and at the same sonic point, 
the energy expression will be the same provided,
$$
\frac{2n_{\rm C}+1}{8}=\frac{2n_{\rm H}+1}{4}=\frac{n_{\rm V}+1}{5} ,
\eqno{(8)}
$$
where, we have used 
$$
(x_c-1)/(x_c-3/5) \sim 1
\eqno{(8a)}
$$
for Model V.

The relations in Eq. (8) are very important. If these relations are 
satisfied, then transonic properties of Model C with polytropic
index $n_{\rm C}$ would be identical to those of Model H with 
index $n_{\rm H}$ and those of Model V with index $n_{\rm V}$ 
respectively.

\section{Shock Invariants and Locations in Different Models}

In between two sonic points, the flow can undergo a
standing shock transition. For an inviscid flow, 
at the shock, a set of conditions are to be satisfied.
These are known as the Rankine-Hugoniot conditions
(Landau \& Lifshitz, 1959). These conditions are different
for different models (C89, C90). For Model V,
the shock conditions are as follows: 
the energy flux  conservation equation, 
$$
\cal E_{+} = E_{-},
\eqno{(9a)}
$$
the pressure balance condition,
$$
W_{+}+\sig_{+} {\vel^{2}_{e}}_{+} = W_{-}+\sig_{-} {\vel^{2}_{e}}_{-}
\eqno{(9b)}
$$
and the baryon flux conservation equation,
$$
{\dot{M}}_{+} ={\dot {M}}_{-}
\eqno{(9c)}
$$
where subscripts ``$-$'' and ``$+$'' refer, respectively, to quantities before 
and after the shock. Here, $W$ and $\sig$ denote the pressure and the density, 
integrated  in the vertical direction (see, e.g., Matsumoto et al. 1984), i.e.,
$$
\sig = \int \limits_{-h}^{h}\rho dz = 2\rho_{e}I_{n}h,
\eqno{(10a)}
$$
and
$$
W = \int \limits_{-h}^{h}P dz = 2P_{e}I_{n+1}h,
\eqno{(10b)}
$$
where, $I_{n} = \frac  {(2^{n}n!)^{2}}{(2n+1)!}$, $n$ being the 
polytropic index as defined previously. 

For Models C and H, the shock conditions  are as follows:
the energy  flux conservation equation,
$$
{\cal E}_+={\cal E}_-
\eqno{(11a)}
$$
the pressure balance condition,
$$
P_{+}+\rho_{+} {\vel^{2}_{e}}_{+} = P_{-}+\rho_{-}
{\vel^{2}_{e}}_{-}
\eqno{(11b)}
$$
and the baryon number conservation equation,
$$
{\dot{M}}_{+} ={\dot {M}}_{-} .
\eqno{(11c)}
$$
The subscripts ``$-$'' and ``$+$'' have the same
interpretation as before. Here, $P$ and $\rho$ denote 
the local pressure and the local density. In the subsequent analysis
we drop the subscript $e$ if no confusion arises in doing so. 

The expressions for the conserved quantities could be combined
to obtain the so-called Mach number relation which must be satisfied 
at the shock. For Model V, we obtain this relation as follows:
We rewrite the energy conservation Eq. (9a), and the pressure balance 
Eq. (9b) in terms of the Mach number $M=\vel/a$ of the flow, 
$$
\frac {1}{2}M^{2}_{+}a^{2}_{+} +\frac {a^{2}_{+}}{\ga -1}
=\frac {1}{2}M^{2}_{-}a^{2}_{-} +\frac {a^{2}_{-}}{\ga -1}
\eqno{(12a)}
$$
$$
{\dot {\cal M_{+}}}=M_{+}a^{\nu^{'}}_{+}f(x_{s})
\eqno{(12b)}
$$
$$
{\dot {\cal M_{-}}}=M_{-}a^{\nu^{'}}_{-}f(x_{s})
\eqno{(12c)}
$$
where, 
$\nu^{'}=\frac {2\ga}{\ga-1}$, and
$$
\frac {a^{\nu}_{+}}{\dot {\cal M}_{+}}{\left( \frac{2}
{3\ga-1}+M^{2}_{+}\right)}=\frac {a^{\nu}_{-}}{\dot 
{\cal M}_{-}}{\left( \frac{2}{3\ga-1}+M^{2}_{-}\right)}
\eqno{(12d)}
$$
where, $\nu = \frac {3\ga -1}{\ga -1}$ and $x_{s}$ is the location of 
the shock. $f(x_s)=x_s^{3/2} (x_s-1)$ is the term in accretion rate
which is explicitly a function of $x$ and is the same both before and
after the shock. From Eqs. 12(a-d) one obtains the following equation 
relating the pre- and post-shock Mach numbers of the flow of Model V at the shock (C89a),
$$
C=\frac {\left[ M_{+}(3\ga-1)+(2/M_{+})\right]^{2}}
{2+(\ga -1)M^{2}_{+}}=\frac {\left[ M_{-}(3\ga-1)+
(2/M_{-})\right]^{2}}{2+(\ga -1)M^{2}_{-}}.
\eqno{(13)}
$$
The constant $C$ is invariant across the shock. The Mach
number of the flow just before and after the shock can 
be written down in terms of $C$ as,
$$
M^{2}_{\mp}=\frac {2(3\ga-1)-C \pm \sqrt {C^{2}-8C\ga}}
{(\ga-1)C-(3\ga-1)^{2}}
\eqno{(14)}
$$
The product of the Mach number is given by,
$$
M_{+}M_{-}=-\frac {2}{\left[ (3\ga-1)^{2}-(\ga-1)C\right]^{1/2}}
\eqno{(15)}
$$

Similarly, one can obtain the Mach-number relations 
and the expression for Mach numbers for the other two models.
The relation between the pre- and the post-shock Mach numbers of the
flow at the shock for Models C and H are given by,
$$
C=\frac {\left[\ga M_{+}+(1/M_{+})\right]^{2}}{2+(\ga -1)
M^{2}_{+}}= \frac {\left[\ga M_{-}+(1/M_{-})\right]^{2}}
{2+(\ga -1)M^{2}_{-}}
\eqno{(16)}
$$

So far, in the literature, analytical shock studies have been carried out in 
models of vertical equilibrium (Das, Chattopadhyay and Chakrabarti, 2001)
by  using the  Mach invariant relations (Eq. 13) when two parameters, 
namely, the specific energy and the specific angular momentum are given.
Presently we carry out the same analysis using Eq. 16 for Models 
H and C respectively and obtained shock locations and parameter 
space boundaries for all the three models. 
Figure 2 compares these results where plots of
specific energy (Y-axis) is given as function of specific angular momentum
(X-axis). Solid boundaries mark regions for which standing shocks form
in different models. Shaded regions are obtained from the analytical
method (Das, Chattopadhyay \& Chakrabarti, 2001) 
and results of these two methods roughly agree.  We note that constant 
height flows occupy much larger region than that of the Conical or vertical equilibrium.

\begin {figure}
\vbox{
\vskip 0.0cm
\hskip 0.0cm
\centerline{
\psfig{figure=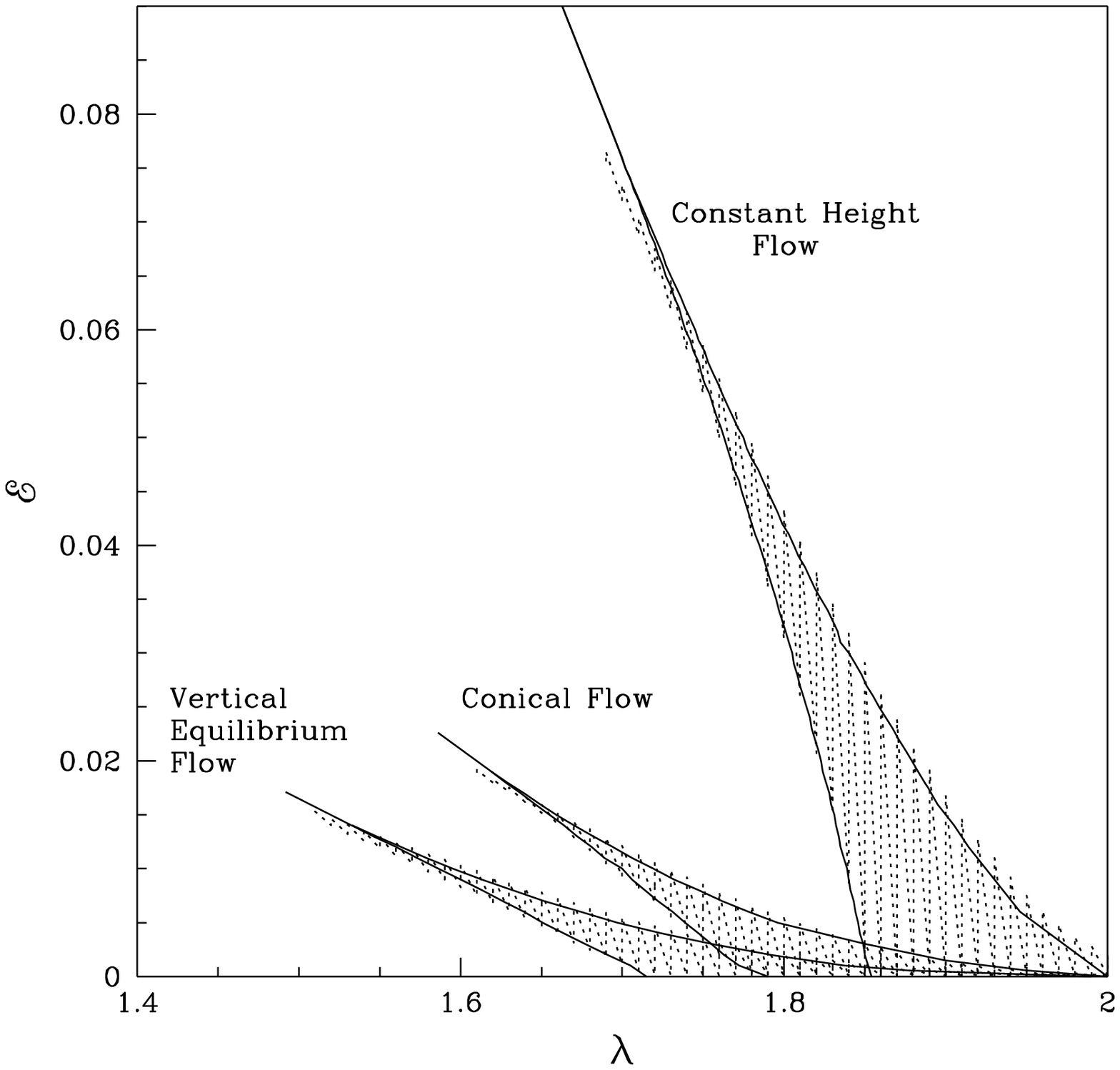,height=10truecm,width=12truecm}}}
\end{figure}
\begin{figure}
\vspace{0.0cm}
\caption[] {Comparison of the parameter space in three different models in which
shocks form. $n_{\rm V}=n_{\rm H}=n_{\rm C}=3$ is chosen throughout.
Solid boundaries are obtained using numerical method and shaded regions are obtained 
using analytical method. }
\end{figure}

\section{Results of Mapping of One Model to Another}

We have already noticed that one could use a relation (Eq. 8) which
maps one model on to another, as far as the transonic properties go.
If, for instance, we choose $n_{\rm H}=3$, we find that $n_{\rm V}=31/4$
and $n_{\rm C}=13/2$ respectively. This means that for a given energy and 
angular momentum a model H flow of polytropic index $3$ would 
have sonic points exactly at the same place as a Model V flow
of polytropic index $31/4$ and Model C flow of polytropic index $13/2$
respectively. Corresponding polytropic exponents are $\gamma_{\rm H}=4/3$, 
$\gamma_{\rm V}=1.129$ and $\gamma_{\rm C}=1.15385$ respectively. 
In physical terms, a relativistic flow of constant height would
have same properties as more or less isothermal flows in a conical flow and
a flow in vertical equilibrium.

What about the shock locations? In Fig. 3, we compare the locations of the
standing shocks around a black in these three models. Solid curves are
for Model V, small circles are for Model H and crosses are for Model C
respectively. The polytropic indices $n_{\rm V}$, $n_{\rm C}$ and $n_{\rm H}$
are as above. We note that though models are different, the 
shock locations are also remarkably close to one another. In Fig. 4,
a comparison of the parameter space is shown once more (cf. Fig. 2). However,
the polytropic indices are chosen as above. Unlike disjoint regions in Fig. 2,
we find that the regions are almost completely over-lapping when 
Eq. (8) was used. This also shows that the Eq. (8) is valid even for the
study of shock waves. We thus believe that generally speaking, the
three models are identical when the Eq. (8) is taken into account.

\begin {figure}
\vbox{
\vskip 0.0cm
\hskip 0.0cm
\centerline{
\psfig{figure=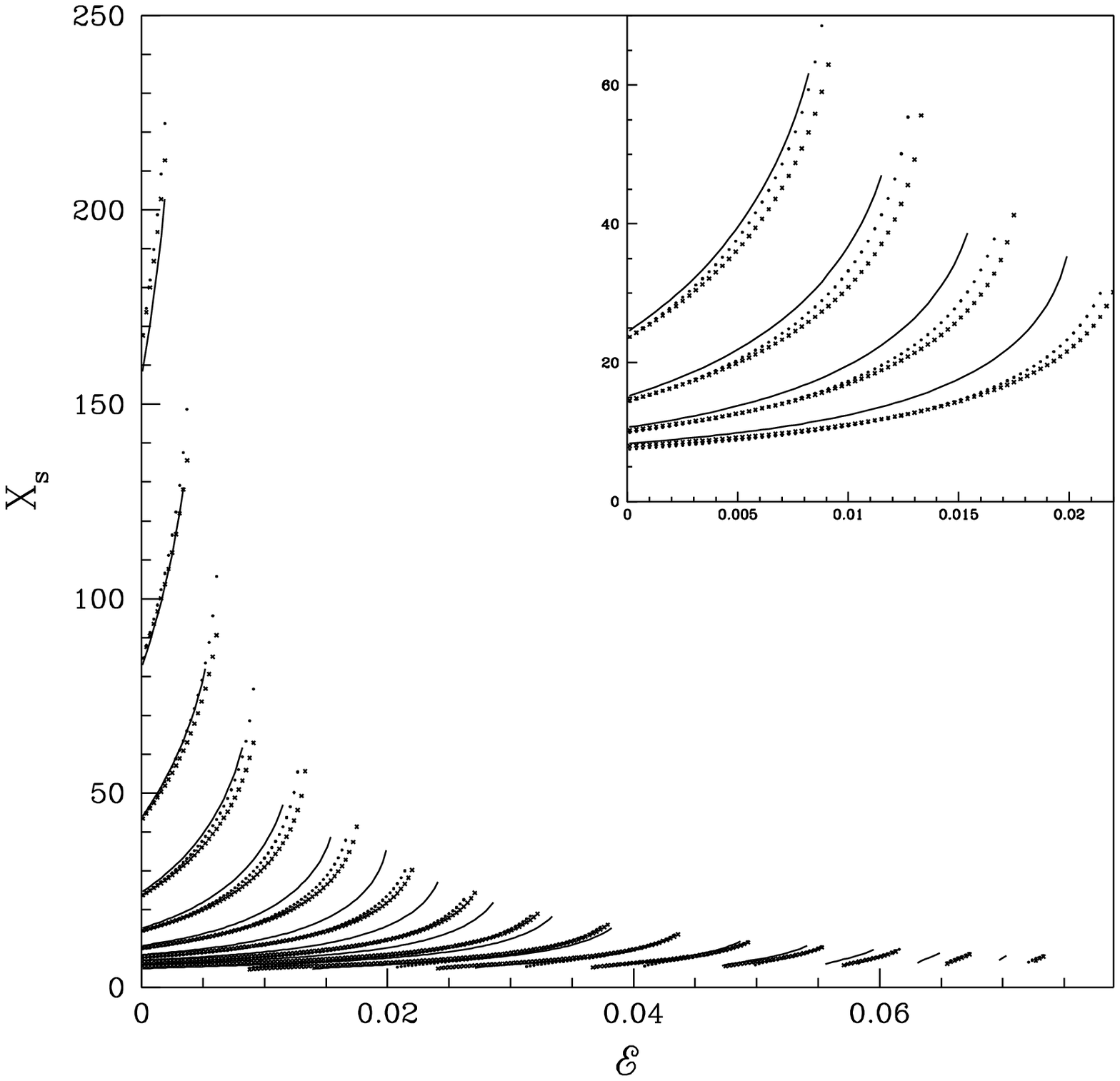,height=10truecm,width=10truecm}}}
\end{figure}
\begin{figure}
\vspace{0.0cm}
\caption[]
{Variations of shock locations (Y-axis) as functions of the specific energy (X-axis)
and angular momenta. The left most curve is drawn for $\lambda=2$ and other 
curves are for decreasing angular momentum with an interval of 
$\delta \lambda=0.02$. Solid curves, filled circles and crosses are drawn 
for Models V, H and C respectively with $n_{\rm V}=31/4$, $n_{\rm C}=13/2$ 
and $n_{\rm H}=3$ which obey Eq. (8).}
\end{figure}

\begin {figure}
\vbox{
\vskip 0.0cm
\hskip 0.0cm
\centerline{
\psfig{figure=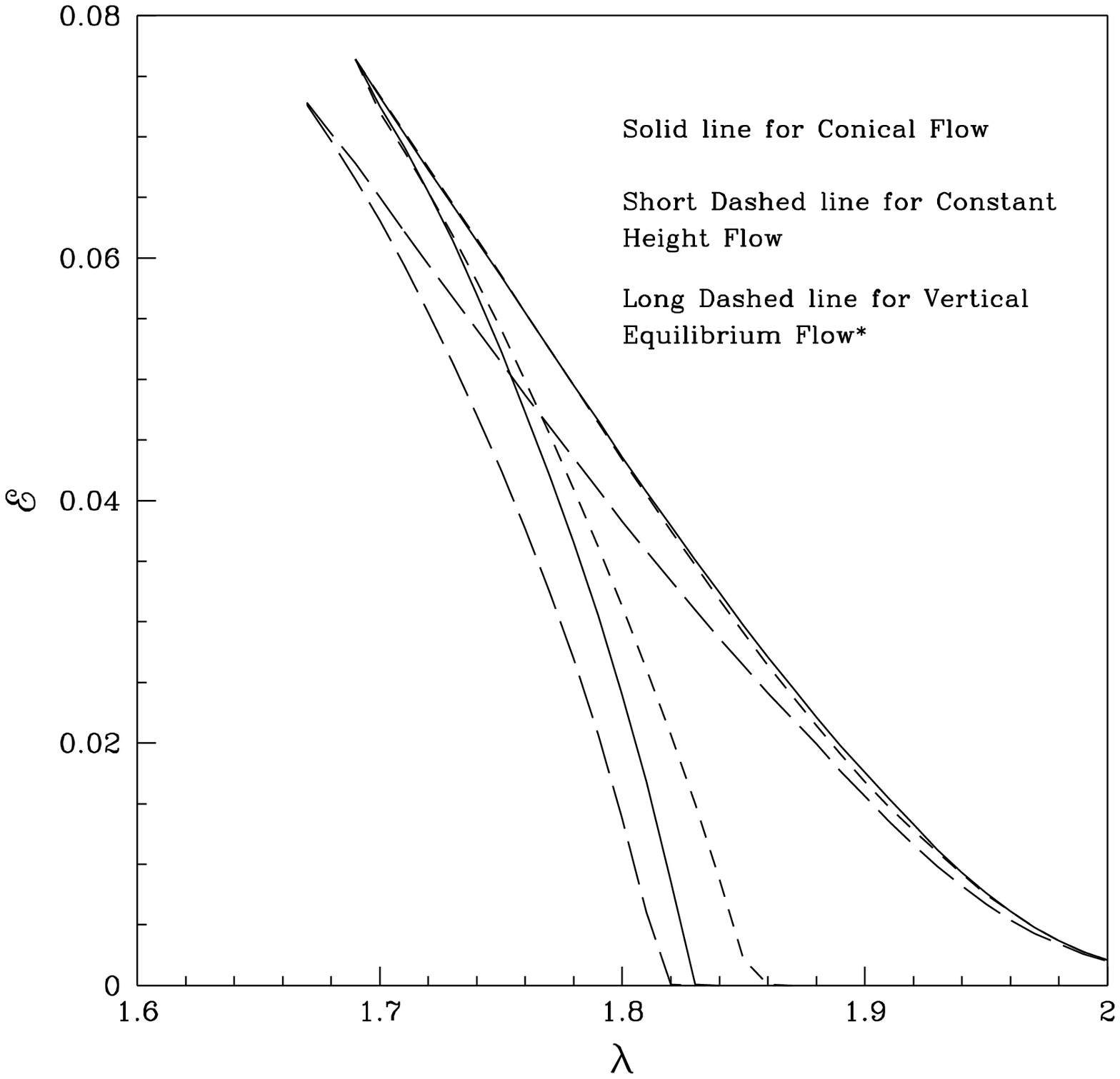,height=10truecm,width=10truecm}}}
\end{figure}
\begin{figure}
\vspace{0.0cm}
\caption[]
{Nature of the boundary of the parameter space for the three models of the
accretion flows. Solid, short-dashed, and long-dashed curves are 
drawn for Models C, H and V respectively with 
polytropic indices $n_{\rm V}=31/4$, $n_{\rm C}=13/2$ and $n_{\rm H}=3$
which obey Eq. (8). The roughly similar parameter space shows that
the mapping of the indices based on the transonic properties
remain roughly the same even when standing shocks are considered. The asterisk mark  
on `Vertical Equilibrium Flow' indicates that condition 8a has been utilized.}
\end{figure}

What could be  possible applications of the pedagogical exercise
we carried out? One could imagine that certain models are easier
to study (say, using numerical simulations) than the others.
For instance, Chakrabarti \& Molteni (1993) and Molteni, Gerardi
and Chakrabarti (1994), Chakrabarti \& Molteni (1995)
studied constant height disks using Smoothed
Particle Hydrodynamics. This was done because a flow in vertical 
equilibrium cannot be forced on a time dependent study. 
However, one could question whether one can draw any conclusion
about the behaviour of flows in vertical equilibrium using a
simulation of constant height. Our present study shows that it does.
Since three models are shown to be identical, running simulation 
for one model would give results for other models in a straight
forward manner. Similarly, study of stability analysis of a
model of constant height may be simpler and  stability of 
one model would imply stability of others.

\section{Concluding Remarks}

In this paper, we discovered a unique relation among 
the polytropic indices of three different models of the 
axisymmetric accretion flows which ensures identical 
transonic properties in the sense that if all these models 
have the same conserved energies and angular momenta, then 
the sonic points also form exactly at the same place. When 
we proceeded further to compute the shock locations, we found 
that even the shocks form roughly at the same places. Apparently, 
disjoint parameter spaces for shock formation with the same
value of polytropic index in three different models 
exhibits considerable overlap when the same unique relation 
(Eq. 8) was used. This shows that the models are virtually
identical in properties and various disk models belong to one 
parameter family.  Our finding has given some insight 
into the relation between the nature of a flow with its
equation of state. It seems that the relativistic 
equation of state in a flow in vertical equilibrium 
behaves similar to a roughly isothermal flow in a disk of constant
height or in a conical flow. It is possible that in the latter
models (Model C and H) the geometric compression is smaller
and hence it is easier to keep them roughly isothermal while conserving
energy as well. 

Though our work has been mainly pedagogical, we believe that it
could have several applications. For instance, linear and non-linear 
stability analysis and time dependent calculations (numerical simulations)
are easier to perform when the disk is of constant 
thickness. Our work indicates that once certain properties regarding stability 
are established in one flow model, they would remain valid in other models as well provided
the relation among the polytropic indices is incorporated.  

The authors greatly acknowledge financial support from Department of Science
and Technology through a Grant (No. SP/S2/K-14/98) with SKC.

{}
\end{document}